\documentclass[prb,twocolumn,showpacs,floatfix]{revtex4}
\usepackage{graphicx}
\usepackage{dcolumn}
\usepackage{bm}

\begin{document}
\title{Space-time evolution of Dirac wave packets}

\author{V.~Ya.~Demikhovskii, G.~M.~Maksimova, A.~A.~Perov, and E.~V.~Frolova}
\email{demi@phys.unn.ru}
\affiliation{Nizhny Novgorod State University,\\
Gagarin Ave., 23, Nizhny Novgorod 603950, Russian Federation}

\date{today}

\begin{abstract}
In this work we study the dynamics of free $3$D relativistic
Gaussian wave packets with different spin polarizations. We
analyze the connection between the symmetry of initial state and
the dynamical characteristics of moving particle. The
corresponding solutions of Dirac equation having different types
of symmetry were evaluated analytically and numerically and after
that the electron probability densities, as well as, the spin
densities were visualized. The average values of velocity of the
packet center and the average spin were calculated analytically,
and the parameters of transient {\it Zitterbewegung} in different
directions were obtained. These results can be useful for the
interpretation of future experiments with trapped ions.
\end{abstract}


\maketitle

\section{Introduction}

The Dirac equation belongs to the most important equations in
modern physics. It predicts the existence of electron spin and
magnetic moment, gives the natural description of the positron
states, describes fine structure of the energy spectrum of
hydrogen-like atoms. The quantized solutions of Dirac equation are
considered to be a natural transition to quantum field theory.
Moreover, the one-particle relativistic quantum mechanics
describes the unexpected electron dynamics including
Schr\"{o}dinger's {\it Zitterbewegung} (ZB)\cite{SchBarut} and
Klein paradox.\cite{Klein} The trembling motion of relativistic
particles or ZB is caused by the interference between positive and
negative energy states which form the electron wave packet. The
frequency of ZB is determined by the gap between two energy bands,
and the amplitude of oscillation of the wave packet center is of
the order of the Compton wavelength.

The results of the first experimental observation of ZB phenomena
were published recently in the paper by Gerrisma {\it et.
al}.\cite{GKZ} For the ZB simulation the experimentalists used a
linear Paul trap where ion motion can be described by
one-dimensional Dirac equation.\cite{Lamata} The authors of
Ref.[3] study the motion of $Ca^+$ ion and determined its position
as a function of time for different initial conditions. As was
shown in Ref.[4] the solution of the $3+1$ Dirac equation can also
be simulated using a single trapped ion with four ionic internal
states. In this case the ion position and momentum are associated
with respective characteristics of $3$D Dirac particle.

The dynamics of relativistic one-dimensional wave packet for the
first time was investigated numerically by Thaller.\cite{Thal} He
visualized the initially localized solutions of single particle
Dirac equation and observed the trembling motion of the wave
packet centers as well as some other phenomena which are caused by
the interference of positive- and negative-energy states. J. Lock
found that ZB oscillations of localized initial states have
transient rather than sustained character.\cite{Lock}  Some
interesting examples of the relativistic dynamics of $3$D electron
wave packets were presented in Ref.[7]. It should be stressed that
ZB oscillations exist only in the one-particle relativistic
theory. Krekora, Su, and Grobe demonstrated analytically and
numerically\cite{Krekora} that quantum field theory does not
permit any {\it Zitterbewegung} of real electrons.

The oscillatory ZB motion of $3$D electron  wave packets in
crystalline solids for the first time was predicted in Ref. [9].
This phenomenon has been considered for 2D electron gas with
Rashba spin-orbit coupling in Ref. [10, 11], in narrow gap
semiconductors in Ref. [12], in carbon nanotubes\cite{Zaw72}, in
single and bilayer graphene\cite{Kat,RusZaw,MDF} and also in
superconductors\cite{Lur}.

In the present work we investigate the relativistic dynamics of
$3$D wave packets. In Sec. II the general properties of symmetry
of Dirac equation solutions which determine the dynamics of wave
packet are analyzed. After that, in Sec.III we consider the
evolution of initial states with different initial symmetry. The
electron probability densities evaluated analytically and
numerically are visualized. The directions of average velocity of
wave packets as well as the phenomenon of high-frequency {\it
Zitterbewegung} for different initial polarization are determined.
In Sec IV the symmetry and structure of spin densities for
relativistic packets are analyzed. Finally, in Sec. V, we conclude
with the discussion of results. Some auxiliary results are found
in Appendixes A and B.

\section{The symmetry of the relativistic wave packets}
In this section we shall first study some general symmetry
properties of the electron wave function which determine the
dynamics of wave packets. So, we start with the famous Dirac
equation for a four-component wave function
$\Psi=(\Psi_1,\Psi_2,\Psi_3,\Psi_4)^T$

$$i\hbar\frac{\partial \Psi}{\partial t}=\hat{H}\Psi,\eqno(1)$$
where $$\hat{H}=-ic\hbar\vec{\alpha}\vec{\nabla}+\beta
mc^2,\eqno(2)$$ $c$ is the light velocity, $m$ is the electron
mass and
$$\vec{\alpha}=\pmatrix{0 & \vec{\sigma} \cr \vec{\sigma} &
0},~~\beta=\pmatrix{I & 0 \cr 0 & -I}.\eqno(3)$$

The four independent free-particle solutions of Eq.(1) for given
momentum $\vec{p}$ and energy $E$, can be written in the form

$$\Psi_{\vec{p},r}(\vec{r},t)=\rm e^{-iEt/\hbar}\varphi_{\vec{p}}(\vec{r})U_r(\vec{p}),~~r=1,2,3,4,\eqno(4)$$
where $\varphi_{\vec{p}}(\vec{r})=\frac{1}{(2\pi \hbar)^{3/2}}\rm
e^{i\vec{p}\vec{r}/\hbar}$, $E=\pm \lambda_{\vec{p}}$,
$\lambda_{\vec{p}}=\sqrt{\vec{p}^2c^2+m^2c^4}$, and $U_r(\vec{p})$
are the free Dirac spinors\cite{Sak}

$$\displaylines{U_1(\vec{p})=N\pmatrix{1 \cr 0 \cr p_3\gamma \cr (p_1+ip_2)\gamma},~
U_2(\vec{p})=N\pmatrix{0 \cr 1 \cr (p_1-ip_2)\gamma \cr
-p_3\gamma}, \cr\hfill
E=\lambda_{\vec{p}}>0,\hfill\llap{(5)}\cr}$$

$$\displaylines{U_3(\vec{p})=N\pmatrix{-p_3\gamma \cr -(p_1+ip_2)\gamma \cr 1 \cr 0},
U_4(\vec{p})=N\pmatrix{-(p_1-ip_2)\gamma \cr p_3\gamma \cr 0 \cr
1}, \cr\hfill E=-\lambda_{\vec{p}}<0,\hfill\llap{(6)}\cr}$$

$$\gamma=\frac{c}{\lambda_{\vec{p}}+mc^2}.\eqno(7)$$
Here we use the normalization condition
$$U_r^+(\vec{p})U_{r^\prime}(\vec{p})=\delta_{rr^\prime},\eqno(8)$$
which means

$$N=\sqrt{(\lambda_{\vec{p}}+mc^2)/2\lambda_{\vec{p}}}.\eqno(9)$$

Let us consider now the symmetry of the solutions of Dirac
equation with respect to the space reflection. Suppose that
initial wave packet is symmetric:
$|\Psi(\vec{r},0)|^2=|\Psi(-\vec{r},0)|^2$. As it known this
symmetry is conserved, i.e.
$|\Psi(\vec{r},t)|^2=|\Psi(-\vec{r},t)|^2$ only if the full
function including the spinor part is the eigenfunction of the
parity operator $\hat{P}=\beta \hat{R}$, where the matrix $\beta$
is determined by Eq.(3) and $\hat{R}$ is the inversion operator
for $\vec{r}$: $\hat{R}\Psi(\vec{r},t)=\Psi(-\vec{r},t)$. Also the
Dirac equation is invariant under the replacement

$$\Psi(x,y,z,t)\rightarrow \Psi^ \prime(\vec{r},t)=\hat{P}_{x,y}\Psi(\vec{r},t),\eqno(10)$$
where $\hat{P}_{x,y}=\Sigma_z\hat{R}_x\hat{R}_y$,
$\hat{R}_x$($\hat{R}_y$) is the inversion operator for $x$($y$)-
component, $\hat{R}_xf(x,y,z)=f(-x,y,z)$ and $\Sigma_z$ is the
corresponding component of spin operator
$\vec{\Sigma}=\pmatrix{\vec{\sigma} & 0 \cr 0 & \vec{\sigma}}$.
So, if the wave function satisfies (in particular at $t=0$) the
relation

$$\hat{P}_{x,y}\Psi(\vec{r},t)=\mp\Psi(\vec{r},t),\eqno(11)$$
then the parity in $x,y$-plane is conserved,

$$|\Psi(x,y,z,t)|^2=|\Psi(-x,-y,z,t)|^2.\eqno(12)$$
Besides, we may determine the reflection transform $z\rightarrow
-z$ in Hilbert space as the operator
$\hat{P}_z=\Sigma_z\beta\hat{R}_z$ which, just as the operators
$\hat{P}$ and $\hat{P}_{x,y}$, commutes with Dirac Hamiltonian
(2). If the initial wave packet has the certain party
$z\rightarrow -z$

$$\hat{P}_z\Psi(\vec{r},0)=\mp \Psi(\vec{r},0),\eqno(13)$$
then $\Psi(\vec{r},t)$ satisfies this equation for all times that
leads to the conservation of the symmetry

$$|\Psi(x,y,z,t)|^2=|\Psi(x,y,-z,t)|^2.\eqno(14)$$
Similarly we may introduce the operators
$\hat{P}_x=\Sigma_x\beta\hat{R}_x$ and
$\hat{P}_y=\Sigma_y\beta\hat{R}_y$ connected with the reflection
transforms $x\rightarrow -x$ and $y\rightarrow -y$
correspondingly.

Thus the symmetry of the solution $\Psi(\vec{r},t)$ with respect
to full ($\vec{r}\rightarrow -\vec{r}$) or partial space symmetry
depends not only on the space symmetry of the initial wave
function, but also on the ratio between its components. This
statement is valid with respect to other types of symmetry. Below
we are interested in the dynamics of the initial wave packet of
the form

$$\Psi(\vec{r},0)=\frac{F(\vec{r})}{\sqrt{\sum\limits_{i=1}^{4}|\varphi_i|^2}}\pmatrix{\varphi_1 \cr \varphi_2 \cr
\varphi_3 \cr \varphi_4},\eqno(15)$$
where $\varphi_i$ are the
complex numbers and the space part $F(\vec{r})$ satisfies the
normalization condition

$$\int|F(\vec{r})|^2d\vec{r}=1.\eqno(16)$$
Specifically we suppose that the probability density at $t=0$
$|\Psi(\vec{r},0)|^2=|F(\vec{r})|^2$ is spherically or axially
symmetric. So let $F(\vec{r})$ has the form

$$F(\vec{r})=F(\rho,z)\rm e^{im\alpha},\eqno(17)$$
where $\rho$, $z$, $\alpha$ are cylindrical coordinates and $m$ is
an integer. The considered wave packet remains an axial
symmetrical (in $x,y$-plane) for all the times  if its spinor part
$\varphi$ is one of the eigenfunctions of operator $\Sigma_z$

$$\Phi_1=\frac{\pmatrix{\varphi_1 & 0 & \varphi_3 & 0}^T}{\sqrt{|\varphi_1|^2+|\varphi_3|^2}}~ or~
\Phi_{-1}=\frac{\pmatrix{0 & \varphi_2 & 0 &
\varphi_4}^T}{\sqrt{|\varphi_2|^2+|\varphi_4|^2}}.\eqno(18)$$
Indeed it is readily to show that the function

$$\Psi(\vec{r},0)=F(\rho,z)\rm e^{im\alpha}\Phi_1,\eqno(19)$$
is the eigenstate of $z$-component of total angular momentum
operator $\hat{I}_z=\hat{l}_z+\frac{1}{2}\Sigma_z$

$$\hat{I}_z\Psi(\vec{r},0)=(m+\frac{1}{2})\Psi(\vec{r},0),\eqno(20)$$
where $\hat{l}_z=-i\partial/\partial\alpha$.

For time $t>0$  the general expression for the wave function is

$$\displaylines{\Psi(\vec{r},t)=(\Psi_1(\rho,\alpha,z,t),\Psi_2(\rho,\alpha,z,t),\cr\hfill
\Psi_3(\rho,\alpha,z,t),\Psi_4(\rho,\alpha,z,t))^T.\hfill\llap{(21)}\cr}$$
But $\hat{I}_z$ is a conserved quantity so that the
$\Psi(\vec{r},t)$ obeys Eq.(20) too. Solving this equation we find
the $\alpha$ -dependence of components $\Psi_i(\rho,\alpha,z,t)$

$$\displaylines{\Psi_1=\rm e^{im\alpha}f_1(\rho,z,t),~\Psi_2=\rm e^{i(m+1)\alpha}f_2(\rho,z,t)\cr\hfill
~\Psi_3=\rm e^{im\alpha}f_3(\rho,z,t),~\Psi_4=\rm
e^{i(m+1)\alpha}f_4(\rho,z,t),\hfill\llap{(22)}\cr}$$ which
immediately leads to conclusion that the probability density is
axially symmetric in $x,y$-plane.

$$|\Psi(\vec{r},t)|^2=\sum\limits_{i=1}^4|f_i(\rho,z,t)|^2.\eqno(23)$$

Note that the initial state $\Psi(\vec{r},0)=F(\rho,z)\rm
e^{im\alpha}\Phi_{-1}$ is the eigenfunction of $\hat{I}_z$ too
with an eigenvalue $m-1/2$. Notice also that as would be shown
below the spherical symmetry of the initial wave packet, Eq. (19)
(with $m=0$ and $F(\rho,z)=F(r)$, $r=\sqrt{\rho^2+z^2}$) turns
into axial one. The above statements concerning the wave packet
symmetry are justified by our analytical and numerical
calculations for different initial conditions.

\section{Dirac wave packets dynamics}

{\bf i) The propagation of cylindrically symmetric wave packet}

Now we describe some peculiarities of the striking kinematics of
three-dimensional relativistic wave packets. As a first example
let us consider the time evolution of the initially localized wave
function of the form

$$\Psi(\vec{r},0)=\frac{F(\vec{r})}{\sqrt{2}}\pmatrix{1 \cr 0 \cr
1 \cr 0}.\eqno(24)$$

As was shown above (see Eq.(18)) for such polarization of the wave
packet the probability density conserves its axial symmetry if the
space part of the wave function (24) $F(\vec{r})$ has the form
(17).

The wave function $\Psi(\vec{r},t)$ can be expanded in plane waves

$$\Psi(\vec{r},t)=\int\varphi_{\vec{p}}(\vec{r})\Psi(\vec{p},t)d\vec{p}.\eqno(25)$$
In the momentum space the components of the bispinor wave function

$$\Psi(\vec{p},t)=(\Psi_1(\vec{p},t),\Psi_2(\vec{p},t),
\Psi_3(\vec{p},t),\Psi_4(\vec{p},t))^T,\eqno(26)$$ are given as a
linear superposition of the positive- and negative-energy
solutions (5) and (6)

$$\Psi(\vec{p},t)=\sum\limits_{r=1}^4C_r(\vec{p})U_r(\vec{p})\rm e^{-iE_rt/\hbar},\eqno(27)$$
where $E_{1,2}=\lambda_{\vec{p}}$ and
$E_{3,4}=-\lambda_{\vec{p}}$. Here $C_r(\vec{p})$ is to be
determined by the Fourier expansion of $\Psi(\vec{r},0)$.
Straightforward calculation using (5), (6) and (24) gives

$$\displaylines{\Psi_1(\vec{p},t)=\frac{f(\vec{p})}{2\sqrt{2}}\Bigg[\rm e^{-i\lambda_{\vec{p}}t/\hbar}(1+
\frac{mc^2+cp_3}{\lambda_{\vec{p}}})+\cr\hfill +\rm
e^{i\lambda_{\vec{p}}t/\hbar}(1-
\frac{mc^2+cp_3}{\lambda_{\vec{p}}})\Bigg],\hfill\llap{(28)}\cr}$$

$$\Psi_2(\vec{p},t)=\Psi_4(\vec{p},t)=-\frac{icf(\vec{p})(p_1+ip_2)}{\lambda_{\vec{p}}\sqrt{2}}\sin
\frac{\lambda_{\vec{p}}t}{\hbar},\eqno(29)$$

$$\displaylines{\Psi_3(\vec{p},t)=\frac{f(\vec{p})}{2\sqrt{2}}\Bigg[\rm
e^{-i\lambda_{\vec{p}}t/\hbar}(1-
\frac{mc^2-cp_3}{\lambda_{\vec{p}}})+\cr\hfill +\rm
e^{i\lambda_{\vec{p}}t/\hbar}(1+
\frac{mc^2-cp_3}{\lambda_{\vec{p}}})\Bigg].\hfill\llap{(30)}\cr}$$
Here $f(\vec{p})$ is the Fourier transform of the function
$F(\vec{r})$. To obtain the components of wave function
$\Psi_i(\vec{r},t)$ we substitute Eqs. (28)-(30) into Eq. (25).
Then for the axially symmetric initial wave packet
$F(\vec{r})=F(\rho,z)$, where $\rho^2=x^2+y^2$, i.e.
$f(\vec{p})=f(p_\bot,p_z)$, $p_\bot^2=p_x^2+p_y^2$, after
integrating over the angular variable we shall have

$$\displaylines{\Psi_1(\rho,z,t)=\frac{1}{4\sqrt{\pi\hbar^3}}\int\limits_{-\infty}^{+\infty}\rm
e^{\frac{ip_zz}{\hbar}}dp_z\int\limits_{0}^{\infty}f(p_\bot,p_z)p_\bot\times
\cr\hfill \times\Bigg[\rm e^{-i\lambda_{\vec{p}}t/\hbar}(1+
\frac{mc^2+cp_3}{\lambda_{\vec{p}}})+\hfill\cr\hfill
+e^{i\lambda_{\vec{p}}t/\hbar}(1-
\frac{mc^2+cp_3}{\lambda_{\vec{p}}})\Bigg]J_0(\frac{p_{\bot}\rho}{\hbar})dp_\bot
,\hfill\llap{(31)}\cr}$$

$$\displaylines{\Psi_2(\rho,\alpha,z,t)=\Psi_4(\rho,\alpha,z,t)=\frac{\rm e^{i\alpha}c}{2\sqrt{\pi\hbar^3}}
\int\limits_{-\infty}^{+\infty} \rm
e^{\frac{ip_zz}{\hbar}}dp_z\times \cr\hfill
 \times \int\limits_{0}^{\infty}f(p_\bot,p_z)\frac{p_\bot^2}{\lambda_{\vec{p}}}J_1(\frac{p_{\bot}\rho}{\hbar})
 \sin \frac{\lambda_{\vec{p}}t}{\hbar}dp_\bot,\hfill\llap{(32)}\cr}$$

 $$\displaylines{\Psi_3(\rho,z,t)=\frac{1}{4\sqrt{\pi\hbar^3}}\int\limits_{-\infty}^{+\infty}\rm
e^{\frac{ip_zz}{\hbar}}dp_z\int\limits_{0}^{\infty}f(p_\bot,p_z)p_\bot\times
\cr\hfill \times\Bigg[\rm e^{-i\lambda_{\vec{p}}t/\hbar}(1-
\frac{mc^2-cp_3}{\lambda_{\vec{p}}})+\hfill\cr\hfill
+e^{i\lambda_{\vec{p}}t/\hbar}(1+
\frac{mc^2-cp_3}{\lambda_{\vec{p}}})\Bigg]J_0(\frac{p_{\bot}\rho}{\hbar})dp_\bot
,\hfill\llap{(33)}\cr}$$ where $J_0(u)$, $J_1(u)$ are Bessel
functions. Using these expressions one may find the dependence of
the electron probability density
$\sum\limits_{i=1}^4|\Psi_i(\vec{r},t)|^2$ on coordinates and time
for the wave packet (24). To illustrate our results in this
Section and everywhere below we choose the function   in the
Gaussian form

$$F(\vec{r})=\frac{1}{d\sqrt{\Delta\pi^3}}\rm e^{-\frac{\rho^2}{2d^2}-\frac{z^2}{2\Delta^2}+ik_0z},\eqno(34)$$
where $d$ and $\Delta$ determine the width of wave packet in the
$x,y$-plane and in $z$ direction correspondingly, and
$\bar{p}_z=\hbar k_0$ is an average momentum of the wave packet.

\begin{figure}
  \centering
  \includegraphics[width=50mm]{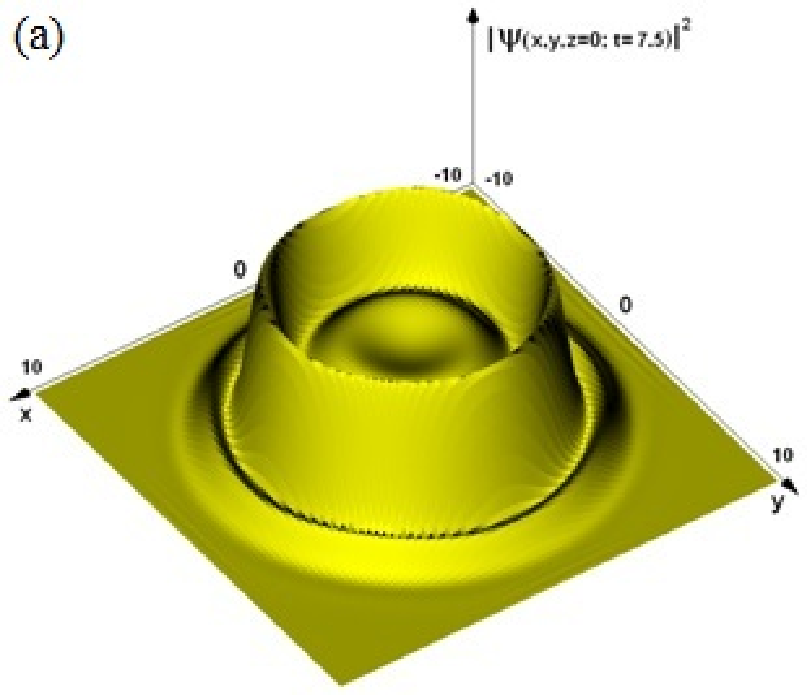}
  \includegraphics[width=50mm]{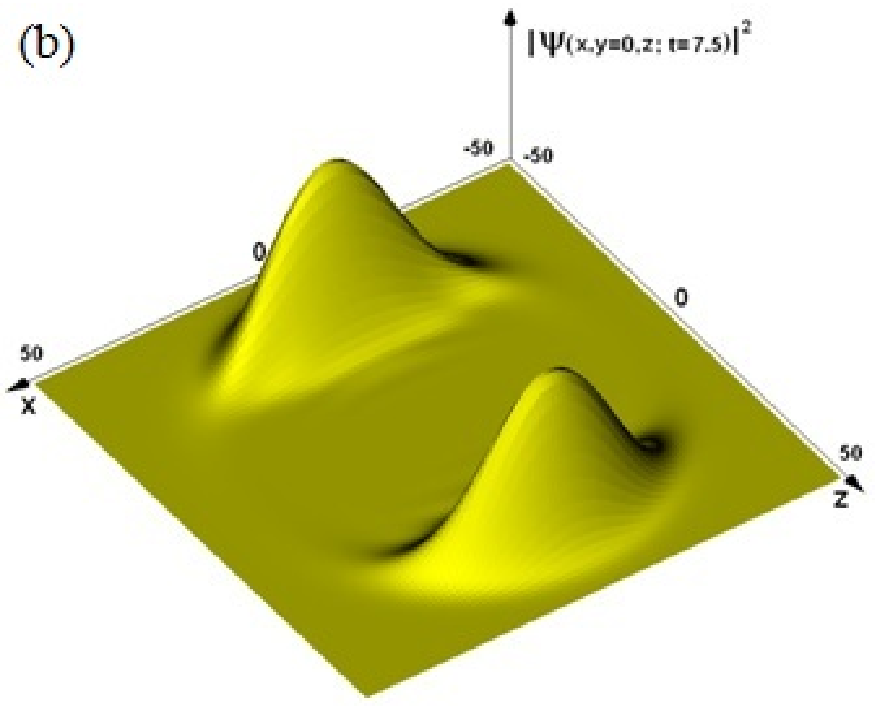}
\caption{(Color online). The electron probability density for the
initial Gaussian packet, Eqs. (24), (34): (a) at $z=0$, $k_0=0$
and $\Delta=5$, $d=1$ at time $t=7.5$; (b) at y=0, $k_0=0$ and
$\Delta=5$, $d=5$ at time $t=7.5$.}\label{Fig1}
\end{figure}

In Fig.1(a) we plot the electron probability density at $z=0$ for
Gaussian wave packet with initial momentum $k_0=0$ and $\Delta=5$,
$d=1$  at time $t=7.5$. Here and below all distances are measured
in units of the Compton wavelength $\lambda_k=\hbar/mc$, the time
is in units of $t_0=\lambda_k/c$. This result was obtained by
using the numerical method ("leap-frog" algorithm) described in
Appendix A. Just as we expected Fig.1(a) demonstrates the axial
symmetry of the considered distribution in $x,y$-plane; the
probability density has a form of cylindrical wave propagating
from the point of origin and having some maxima.

To analyze the motion of the packet we have to find the average
value of velocity of the packet center. In the momentum
representation

$$\bar{V}_i(t)=c\int\Psi^+(\vec{p},t)\alpha_i\Psi(\vec{p},t)d\vec{p},~~i=1,2,3.\eqno(35)$$
Substituting Eqs.(28)-(30) into Eq.(35) and using the expressions
for matrices  $\alpha_i$ (Eq.(3)) we obtain

$$\bar{V}_x(t)=c\int d\vec{p}|f(\vec{p})|^2\Bigg[\frac{c^2p_1p_3}{\lambda_{\vec{p}}^2}(1-\cos\frac{2\lambda_{\vec{p}}t}{\hbar})+
\frac{cp_2}{\lambda_{\vec{p}}}\sin\frac{2\lambda_{\vec{p}}t}{\hbar}\Bigg],\eqno(36)$$

$$\bar{V}_y(t)=c\int
d\vec{p}|f(\vec{p})|^2\Bigg[\frac{c^2p_2p_3}{\lambda_{\vec{p}}^2}(1-\cos\frac{2\lambda_{\vec{p}}t}{\hbar})-
\frac{cp_1}{\lambda_{\vec{p}}}\sin\frac{2\lambda_{\vec{p}}t}{\hbar}\Bigg],\eqno(37)$$

$$\bar{V}_z(t)=c\int d\vec{p}|f(\vec{p})|^2\Bigg[\frac{c^2p_3^2}{\lambda_{\vec{p}}^2}+
(1-\frac{c^2p_3^2}{\lambda_{\vec{p}}^2})\cos\frac{2\lambda_{\vec{p}}t}{\hbar}
\Bigg],\eqno(38)$$ Since the Fourier transform $f(\vec{p})$ of
Gaussian wave packet (34) has an axial symmetry in $p_x,p_y$-
plane

$$f(\vec{p})=\frac{d\sqrt{d}}{(\hbar\sqrt{\pi})^{3/2}}\rm e^{-\frac{p_\bot^2d^2}{2\hbar^2}-\frac{(p_z-\hbar k_0)^2\Delta^2}{2\hbar^2}},\eqno(39)$$
the components of velocity $\bar{V}_x=\bar{V}_y=0$ as it follows
from Eqs.(36),(38). Otherwise, owing to the axial symmetry of
spacial distribution of the electron density, the average
coordinates of packet $\bar{x}=\bar{y}=0$. As a result, mean
components of velocity in $x,y$-plane are equal to zero.

The motion of the packet center in $z$-direction experiences rapid
oscillations commonly known as {\it Zitterbewegung} (the second
term in square brackets in Eq.(38)). Besides, the wave packet
displaces slowly with constant velocity $\bar{V}_{z0}$ (the first
term in square brackets in Eq.(38); see also Eq.(B.8)) even if its
momentum $\bar{p}_z=\hbar k_0=0$.

The existence of the constant component of average velocity
$\bar{V}_{z0}$ in this case can be understood from the relation
between velocity and momentum depending on the sign of energy: for
the wave packet consisting of the states with positive (negative)
energy a positive momentum $p_z$ corresponds to a positive
(negative) velocity $V_z$. Let us represent the time-independent
probability density in the momentum space $|\Psi(\vec{p})|^2$ as a
superposition of the positive energy part $\Psi_+(\vec{p})$ and
the negative energy part $\Psi_-(\vec{p})$

$$|\Psi(\vec{p})|^2=|\Psi_+(\vec{p})|^2+|\Psi_-(\vec{p})|^2,\eqno(40)$$
where
$|\Psi_{\pm}(\vec{p})|^2=\sum\limits_{i=1}^4|\Psi_{i\pm}(\vec{p},t)|^2$,
$\Psi_{i\pm}(\vec{p},t)\propto\rm e^{\mp
i\frac{\lambda_{p}t}{\hbar}}$. Using Eqs.(28)-(30) one may readily
show that

$$|\Psi_{\pm}(\vec{p})|^2=\frac{|f(\vec{p})|^2}{2}(1\pm\frac{cp_z}{\lambda_{\vec{p}}}),\eqno(41)$$

Fig.2 shows the dependence
$W_{\pm}(p_z)=2\pi\int\limits_{0}^\infty|\Psi_\pm(\vec{p})|^2p_\bot
dp_\bot$ (in arbitrary units) on $p_z$ (in units of $mc$) for the
Gaussian initial wave packet, Eqs. (24), (39) for $k_0=0$ (a) and
for $k_0=1$ (in units of $mc/\hbar$) (b).

\begin{figure}
  \centering
  \includegraphics[width=50mm]{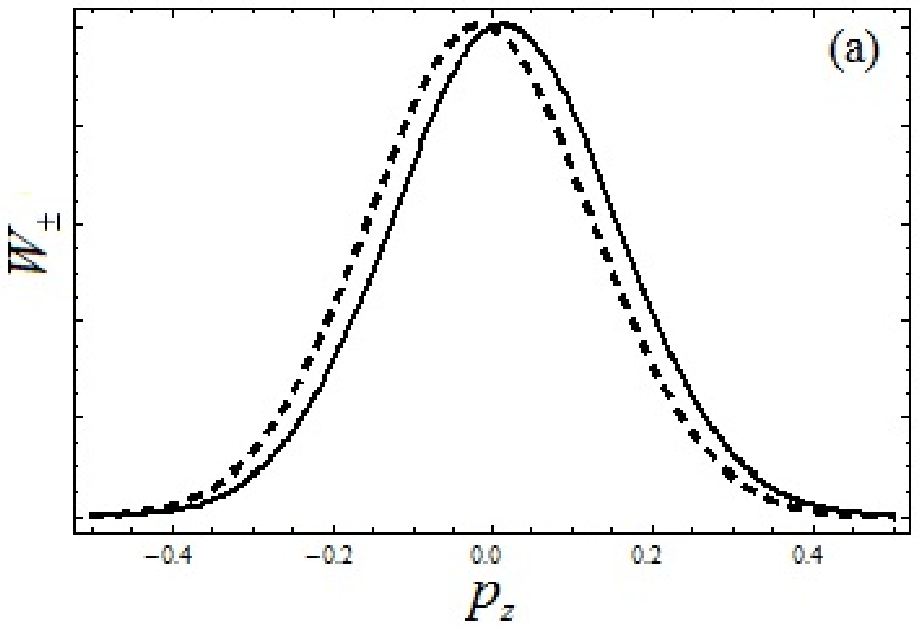}
  \includegraphics[width=50mm]{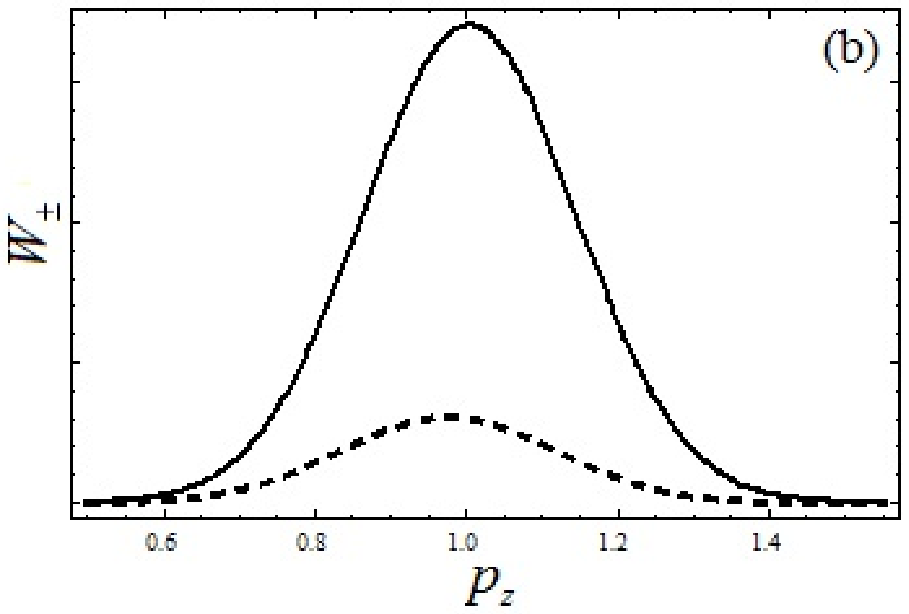}
\caption{(Color online). The dependencies $W_{\pm}(p_z)$ of
positive (solid line) and negative (dashed line) energy parts for
Gaussian initial wave packet, Eqs. (24), (39) for $k_0=0$,
$\Delta=5$, $d=1$ (a), and for  $k_0=1$, $\Delta=5$, $d=5$
(b).}\label{Fig2}
\end{figure}

We see that for the case $k_0=0$ the momentum distribution for
positive- and negative- energy components is shifted towards
positive and negative momentum $p_z$, respectively. But in both
cases such dependence leads to the motion of each parts (and the
whole wave packet) with a positive velocity along $z$-axis. We
also see in Fig. 2(a) that the function $W_+(p_z)$ has essential
overlap with the function $W_-(p_z)$ that is a necessary condition
for existence of {\it Zitterbewegung} of the packet center (see
Eq.(38)). The value of constant component of velocity
$\bar{V}_{z0}$ depends on the ratio between the initial width of
the wave packet and the Compton wavelength. In the limiting case
$d>>1$ $\bar{V}_{z0}$ is much less than the light velocity:
$\bar{V}_{z0}=\frac{c}{2}(\frac{1}{d})^2$. In particular, for the
symmetrical wave packet ($k_0=0$) of width $d=\Delta=5$
$\bar{V}_{z0}\approx0.02c$.

Let now the initial wave function $F(\vec{r})$ describes the wave
packet moving along $z$-axis with average momentum
$\bar{p}_z=\hbar k_0$. Then the distribution of the full electron
density in $x,y$-plane is similar to one shown in Fig.1(a).
However the essential difference appears in the character of
evolution of the wave packet in $x,z$ (or $y,z$)-plane (Fig.
1(b)). The dependencies $W_\pm (p_z)$ for the states with
positive- and negative- energy parts for this case is shown in
Fig. 2(b). We see that both components consist of positive
momentum $p_z$. For the smaller negative-energy parts, this
corresponds to the motion with negative velocity along $z$-axis.
Thus in the position space the initial wave packet splits into two
packets propagating in the opposite directions along $z$ axis.

{\bf ii) Asymmetrical wave packet evolution}

We next consider another example of the initial spin polarization
of the electron

$$\Psi(\vec{r},0)=\frac{F(\vec{r})}{\sqrt{2}}\pmatrix{1 \cr 0 \cr
0 \cr 1},\eqno(42)$$ where as before the function $F(\vec{r})$  is
determined by Eq.(34).

Note that the example under review is invariant with respect to
the reflection transformation $z\rightarrow -z$, i.e. the
expression (42) satisfies Eq.(13). As was shown above this means
that the probability density is an even function of $z$ at all
times. Performing the same kind of calculations as for symmetrical
wave packet we find the components of the initial wave function
(42) at $t>0$ in the momentum space.

$$\displaylines{\Psi_1(\vec{p},t)=\frac{f(\vec{p})}{2\sqrt{2}}\Bigg[e^{-i\lambda_{\vec{p}}t/\hbar}(1+
\frac{mc^2+c(p_1-ip_2)}{\lambda_{\vec{p}}})+\cr\hfill +\rm
e^{i\lambda_{\vec{p}}t/\hbar}(1-
\frac{mc^2+c(p_1-ip_2)}{\lambda_{\vec{p}}})\Bigg],\hfill\llap{(43)}\cr}$$

$$\Psi_2(\vec{p},t)=-\Psi_3(\vec{p},t)=\frac{icf(\vec{p})p_3}{\lambda_{\vec{p}}\sqrt{2}}\sin
\frac{\lambda_{\vec{p}}t}{\hbar},\eqno(44)$$

$$\displaylines{\Psi_4(\vec{p},t)=\frac{f(\vec{p})}{2\sqrt{2}}\Bigg[\rm
e^{-i\lambda_{\vec{p}}t/\hbar}(1-
\frac{mc^2-c(p_1+ip_2)}{\lambda_{\vec{p}}})+\cr\hfill +\rm
e^{i\lambda_{\vec{p}}t/\hbar}(1+
\frac{mc^2-c(p_1+ip_2)}{\lambda_{\vec{p}}})\Bigg].\hfill\llap{(45)}\cr}$$
The components of $\Psi(\vec{r},t)$ can be obtained directly by
the Fourier transform of Eqs.(43)-(45).

$$\displaylines{\Psi_1(\rho,\alpha,t)=\frac{1}{2\sqrt{\pi\hbar^3}}\int\limits_{-\infty}^{+\infty}\rm
e^{\frac{ip_zz}{\hbar}}dp_z\int\limits_{0}^{\infty}f(p_\bot,p_z)p_\bot\times
\cr\hfill
\times\Bigg[\cos\frac{\lambda_{\vec{p}}t}{\hbar}J_0(\frac{p_\bot
\rho}{\hbar})+\frac{cp_\bot}{\lambda_{\vec{p}}}\sin\frac{\lambda_{\vec{p}}t}{\hbar}\rm
e^{-i\alpha}J_1(\frac{p_\bot \rho}{\hbar})-\hfill\cr\hfill -
\frac{imc^2}{\lambda_{\vec{p}}}\sin\frac{\lambda_{\vec{p}}t}{\hbar}J_0(\frac{p_{\bot}\rho}{\hbar})\Bigg]dp_\bot
,\hfill\llap{(46)}\cr}$$

$$\displaylines{\Psi_2(\rho,z,t)=-\Psi_3(\rho,z,t)=\frac{ic}{2\sqrt{\pi\hbar^3}}
\int\limits_{-\infty}^{+\infty}p_z \rm
e^{\frac{ip_zz}{\hbar}}dp_z\times \cr\hfill
 \times \int\limits_{0}^{\infty}f(p_\bot,p_z)\frac{p_\bot}{\lambda_{\vec{p}}}\sin \frac{\lambda_{\vec{p}}t}{\hbar}J_0(\frac{p_{\bot}\rho}{\hbar})
 dp_\bot,\hfill\llap{(47)}\cr}$$

$$\displaylines{\Psi_4(\rho,\alpha,t)=\frac{1}{2\sqrt{\pi\hbar^3}}\int\limits_{-\infty}^{+\infty}\rm
e^{\frac{ip_zz}{\hbar}}dp_z\int\limits_{0}^{\infty}f(p_\bot,p_z)p_\bot\times
\cr\hfill \times
\Bigg[\cos\frac{\lambda_{\vec{p}}t}{\hbar}J_0(\frac{p_\bot
\rho}{\hbar})+\frac{cp_\bot}{\lambda_{\vec{p}}}\sin\frac{\lambda_{\vec{p}}t}{\hbar}\rm
e^{i\alpha}J_1(\frac{p_\bot \rho}{\hbar})+\hfill\cr\hfill +
\frac{imc^2}{\lambda_{\vec{p}}}\sin\frac{\lambda_{\vec{p}}t}{\hbar}J_0(\frac{p_{\bot}\rho}{\hbar})\Bigg]dp_\bot
.\hfill\llap{(48)}\cr}$$ Using these expressions one may find the
full electron density $|\Psi(\vec{r},t)|^2$. Figure 3(a) shows the
corresponding probability density distribution in the plane $z=0$
at time $t=2\pi$. The parameters of wave packet are: $k_0=0$ and
$\Delta=5$, $d=1$. Comparing Figs. 1(a) and 3(a) we see that the
change of initial spin polarization leads to the fact that the
wave packet being axially symmetric in $x,y$-plane at $t=0$ loses
its symmetry at $t>0$.

\begin{figure}
  \centering
  \includegraphics[width=50mm]{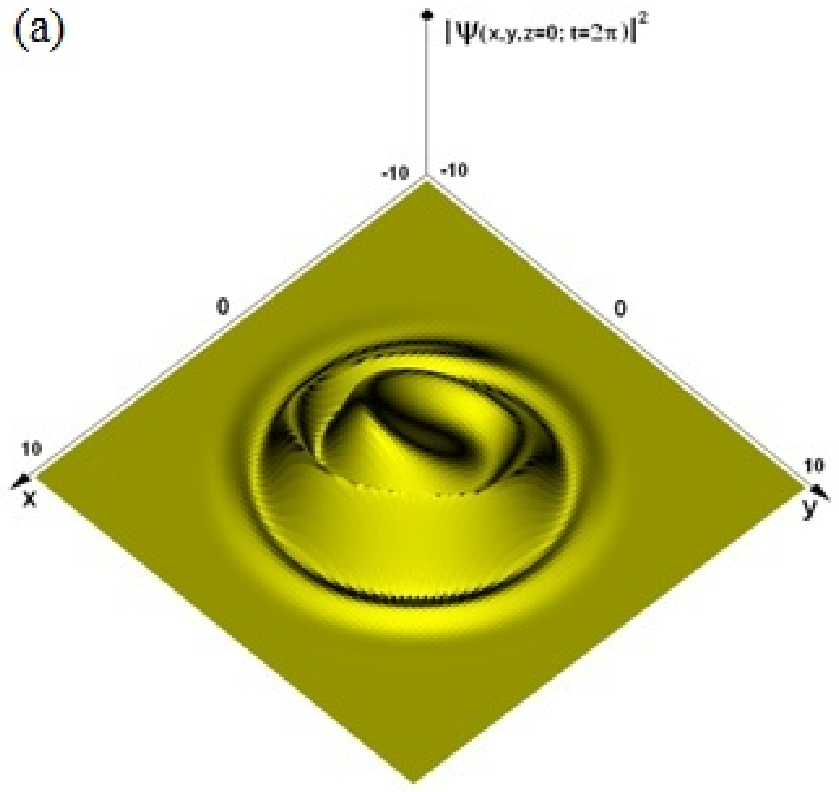}
  \includegraphics[width=50mm]{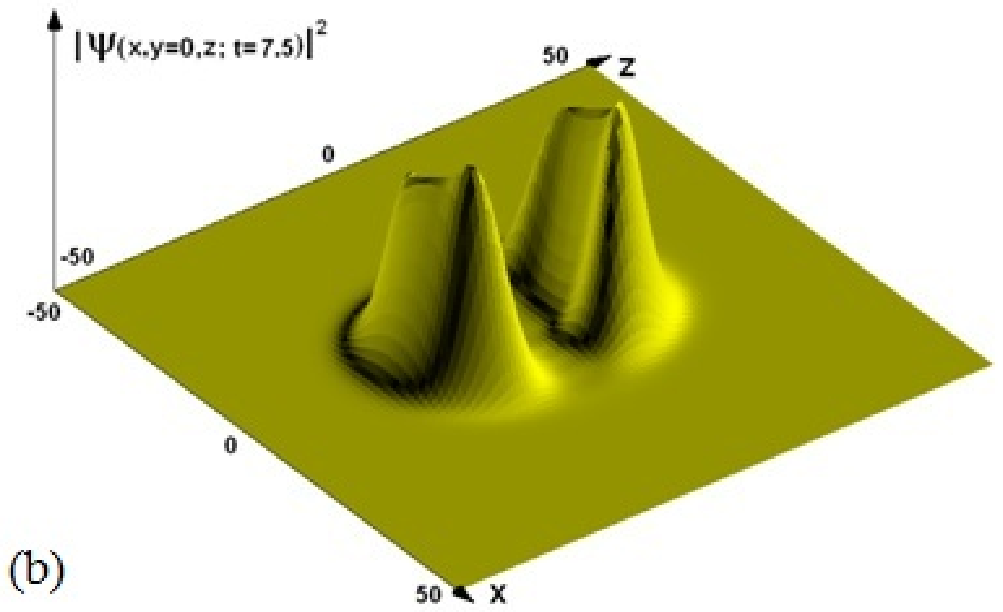}
\caption{(Color online). The electron probability density for the
initial Gaussian packet, Eqs. (42), (34): (a) at $z=0$ with
$k_0=0$ and $\Delta=5$, $d=1$ at time $t=2\pi$; (b) at y=0,
$k_0=1$ and $\Delta=5$, $d=2.5$ at time $t=7.5$.}\label{Fig3}
\end{figure}

The kinematics of the wave packet can be characterized by the
average velocity of its center with components

$$\bar{V}_x(t)=c\int d\vec{p}|f(\vec{p})|^2\Bigg[\frac{c^2p_1^2}{\lambda_{\vec{p}}^2}+
(1-\frac{c^2p_1^2}{\lambda_{\vec{p}}^2})\cos\frac{2\lambda_{\vec{p}}t}{\hbar}\Bigg],\eqno(49)$$

$$\bar{V}_y(t)=c\int d\vec{p}|f(\vec{p})|^2\Bigg[\frac{c^2p_1p_2}{\lambda_{\vec{p}}^2}(1-\cos\frac{2\lambda_{\vec{p}}t}{\hbar})+
\frac{mc^2}{\lambda_{\vec{p}}}\sin\frac{2\lambda_{\vec{p}}t}{\hbar}\Bigg],\eqno(50)$$

$$\bar{V}_z(t)=c\int d\vec{p}|f(\vec{p})|^2\frac{c^2p_1p_3}{\lambda_{\vec{p}}^2}\Bigg[
1-\cos\frac{2\lambda_{\vec{p}}t}{\hbar}\Bigg].\eqno(51)$$ As in
the previous case the wave packet center drifts with constant
velocity (the first term in square brackets in Eq.(49)), but now
it is directed along $x$ axis. Besides, one can see from Eqs. (49)
and (50) that the packet center performs damped oscillations (Fig.
4) along $x$ and $y$ directions. We also see from Eq.(51) that
$\bar{V}_z=0$ that is the result of the symmetry under the
replacement $z\rightarrow -z$, Eq.(14).

\begin{figure}
  \centering
  \includegraphics[width=50mm]{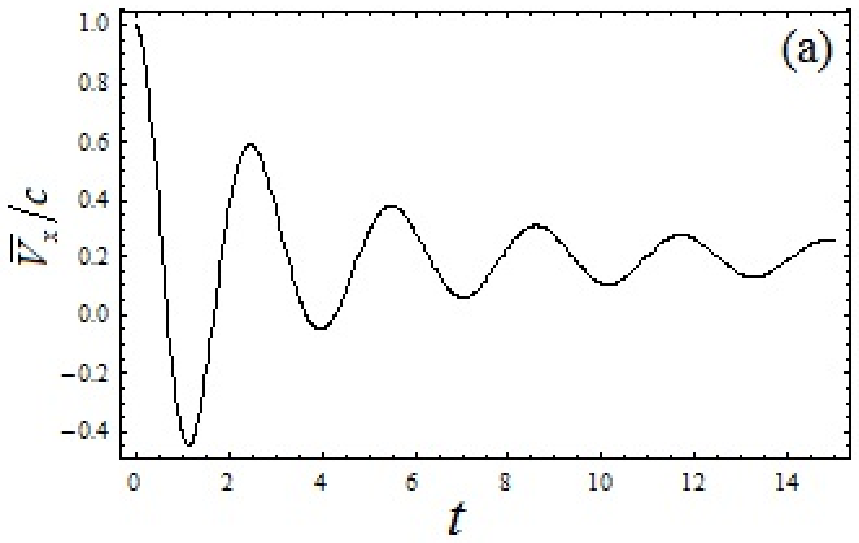}
  \includegraphics[width=50mm]{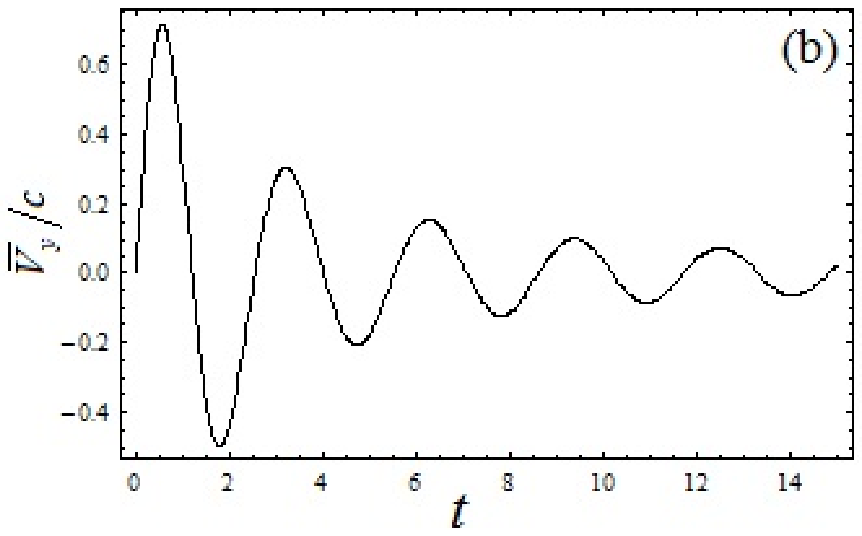}
\caption{(Color online). The average projections of velocities
$\bar{V}_x/c$ (a) and $\bar{V}_y/c$ (b) versus time $t$ (in units
of $\lambda_k/c$) for Gaussian initial wave packet, Eqs. (42),
(34) with $k_0=0$ and $\Delta=5$, $d=1$.}\label{Fig4}
\end{figure}

We now consider the behavior of the initial wave packet with
nonzero momentum $p_z=\hbar k_0$. Obviously such initial state is
not an eigenfunction of the parity operator $\hat{P}_z$.
Nevertheless, the probability density remain to be a symmetrical
function relatively to the reflection transform $z\rightarrow -z$.
Indeed, one may check that the Dirac equation (1) is invariant
under the transformation

$$\Psi(\vec{r},t)\rightarrow \Psi'(x,y,z,t)=\alpha_x \Psi^*(x,y,-z,t).\eqno(52)$$
So that if the initial wave function satisfies the equation

$$\alpha_x \Psi^*(x,y,-z,0)=\Psi(x,y,z,0),\eqno(53)$$
then, as one can check using Eqs.(46)-(48), this relationship is
valid at $t>0$ and consequently Eq. (14) holds.

The character of the motion of the packet center in $x,y$-plane is
similar to the case $k_0=0$. Namely, the center of the wave packet
drifts along $x$-direction with constant component of velocity and
oscillates along $x$ and $y$ axis ({\it Zitterbewegung}). As it
known the ZB is significant if the subpackets with positive and
negative energy have overlap in the position space. But if the
initial average momentum of the wave packet is nonzero, both
subpackets move with the opposite velocities along $z$ axis that
leads to their spatial separation. So, the amplitude of the ZB
decreases more rapidly than for the case $k_0=0$ (compare Fig. 4
and Fig. 5). Notice that this result is valid for the narrow
enough wave packets with width $d\leq 1$ and $dk_0\ll 1$ (for very
large $d$ the ZB oscillations are almost undamped). The constant
component of packet center velocity $\bar{V}_{x0}$ also depends on
the initial momentum $\hbar k_0$. In fact, as it follows from Eq.
(49) it decreases as $k_0$ increases.

\begin{figure}
  \centering
  \includegraphics[width=50mm]{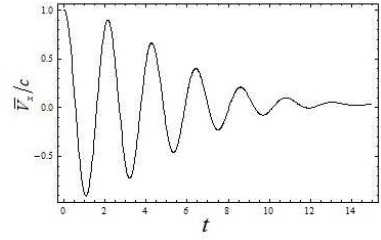}
\caption{(Color online). The average projection of velocity
$\bar{V}_x/c$(a) versus time $t$ (in units of $\lambda_k/c$) for
Gaussian initial wave packet, Eqs. (42), (34) with $k_0=1$ and
$\Delta=5$, $d=2.5$.}\label{Fig4}
\end{figure}

\section{Spin dynamics}

At present we consider the average value of the spin operator
$\vec{\Sigma}$ for Dirac particle

$$\bar{\Sigma}_\mu(t)=\int d\vec{r}\Sigma_\mu(\vec{r},t)=\int d\vec{r}\sum\limits_{i=1}^{4}\Psi_i^+(\vec{r},t)\Sigma_\mu\Psi_i(\vec{r},t),\eqno(54a)$$
or in the momentum representation
$$\bar{\Sigma}_\mu(t)=\int d\vec{p}\sum\Psi_i^+(\vec{p},t)\Sigma_\mu
\Psi_i(\vec{p},t).\eqno(54b)$$

One can verify that the spin densities $\Sigma_\mu(\vec{r},t)$ for
the arbitrary wave function with components
$\Psi_i=\Psi_i(\vec{r},t)$ are

$$\Sigma_x(\vec{r},t)=2{\rm Re}(\Psi_1^*\Psi_2+\Psi_3^*\Psi_4),\eqno(55)$$

$$\Sigma_y(\vec{r},t)=2{\rm Im}(\Psi_1^*\Psi_2+\Psi_3^*\Psi_4),\eqno(56)$$

$$\Sigma_z(\vec{r},t)=|\Psi_1|^2-|\Psi_2|^2+|\Psi_3|^2-|\Psi_4|^.\eqno(57)$$

{\bf i) Cylindrically symmetric wave packet}

The form of Eqs.(55)-(57) and the expressions (31)-(33) for the
components of wave function show that for an axially symmetric
wave packet, Eq.(24) the component of spin density $\Sigma_z$ is
an axially symmetric too both for $k_0=0$ and $k_0\neq 0$ (Fig.
6(c)). One can also verify that $\Sigma_x$ and $\Sigma_y$
components can be written in the form

\begin{figure}
  \centering
  \includegraphics[width=45mm]{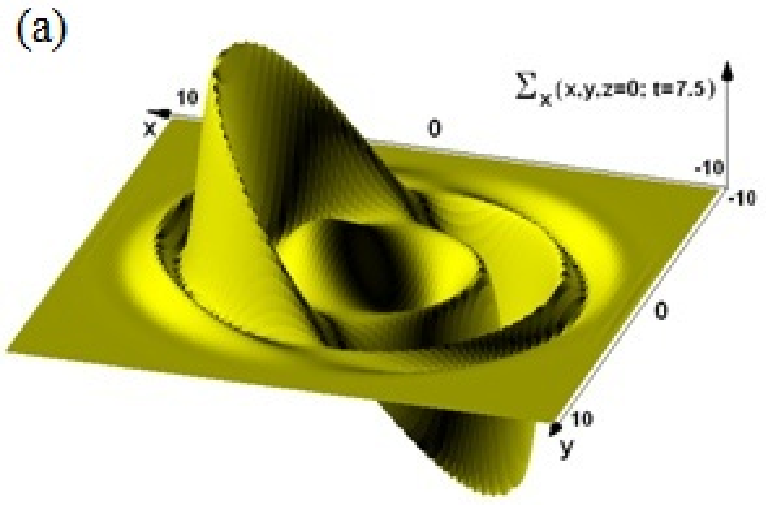}
  \includegraphics[width=45mm]{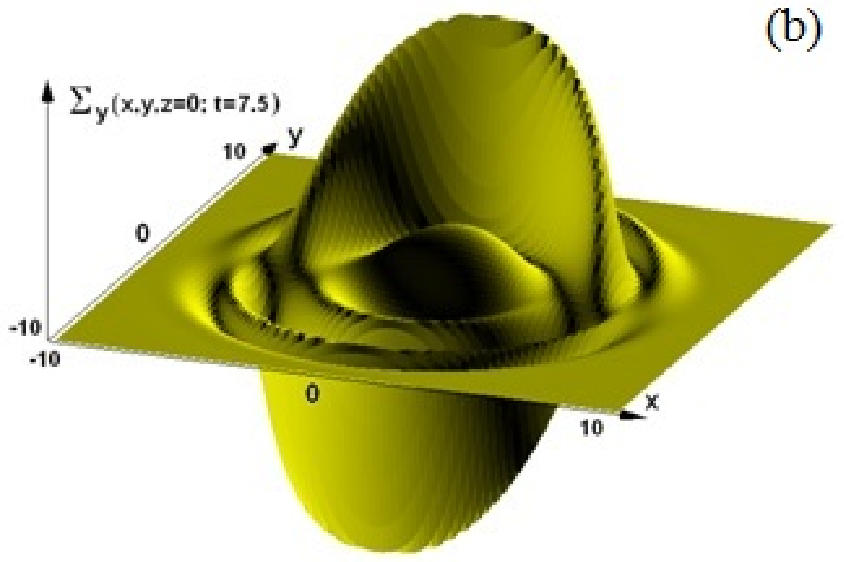}
  \includegraphics[width=50mm]{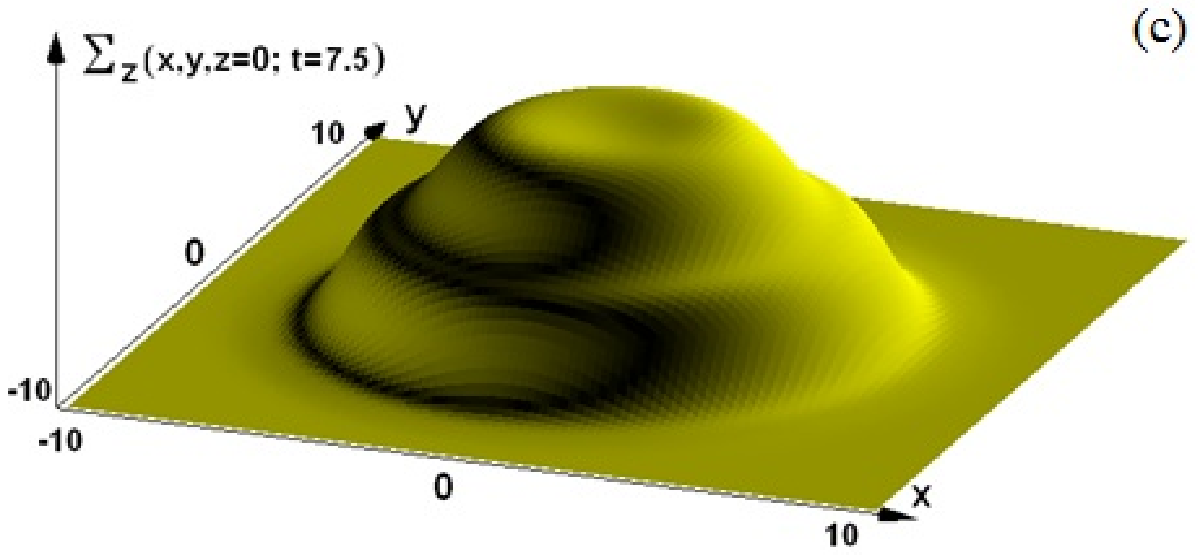}
\caption{(Color online). The distributions of the components of
spin density $\Sigma_x(x,y,0,t)$, $\Sigma_y(x,y,0,t)$,
$\Sigma_z(x,y,0,t)$ for initial Gaussian packet, Eqs. (24), (34)
with $k_0=0$ and $\Delta=5$, $d=1$ at time $t=7.5$.}\label{Fig6}
\end{figure}

$$\Sigma_x=x\cdot g(\rho,z,t),~~~\Sigma_y=y\cdot g(\rho,z,t),\eqno(58)$$
where the expression for the function $g(\rho,z,t)$ is enough
cumbrous and will not be presented here. Thus the $x,y$-plane
projection of vector $\vec{\Sigma}(\vec{r},t)$ is directed along
 the unit vector of cylindrical system $\vec{e}_\rho$. The spin
densities $\Sigma_x(x,y,0,t)$ and $\Sigma_y(x,y,0,t)$ are
represented in Fig. 6(a),(b) for the Gaussian wave packet with the
average momentum $\bar{p}_z=\hbar k_0=0$. We see that
$\Sigma_x(x,y,0,t)$ ($\Sigma_y(x,y,0,t)$) is the antisymmetrical
function of $x$($y$) and the spin density $\Sigma_z(x,y,0,t)$
conserves its axial symmetry.

The average values of spin operators for this polarization can be
found by using Eqs.(28)-(30) for wave function in the momentum
representation and previous definition, Eq.(54b).

$$\bar{\Sigma}_x(t)=\int\frac{|f(\vec{p})|^2}{\lambda_{\vec{p}}^2}\Bigg[cp_2\lambda_{\vec{p}}\sin\frac{2\lambda_{\vec{p}}t}{\hbar}+
c^2p_3p_1(1-\cos\frac{2\lambda_{\vec{p}}t}{\hbar})\Bigg]d\vec{p},\eqno(59)$$

$$\bar{\Sigma}_y(t)=\int\frac{|f(\vec{p})|^2}{\lambda_{\vec{p}}^2}\Bigg[-cp_1\lambda_{\vec{p}}\sin\frac{2\lambda_{\vec{p}}t}{\hbar}+
c^2p_3p_2(1-\cos\frac{2\lambda_{\vec{p}}t}{\hbar})\Bigg]d\vec{p},\eqno(60)$$

$$\bar{\Sigma}_z(t)=\int\frac{|f(\vec{p})|^2}{\lambda_{\vec{p}}^2}\Bigg[(\lambda_{\vec{p}}^2-c^2p_\bot^2)+c^2p_\bot^2
\cos\frac{2\lambda_{\vec{p}}t}{\hbar}\Bigg]d\vec{p}.\eqno(61)$$ As
it follows from these relations only $\bar{\Sigma}_z(t)$ is not
equal to zero for the considered wave packet. The first term in
square brackets of the last formula corresponds to the constant
component of $\bar{\Sigma}_z(t)$ and the second one describes the
typical transient {\it Zitterbewegung}.

{\bf ii) Asymmetrical wave packet}

One can easy verify that the second example, Eq.(42) considered in
our work corresponds to the initial spin density
$\vec{\Sigma}(\vec{r},0)=0$. Really, as it follows from
Eqs.(46)-(48) the components of wave function for the packet with
$k_0=0$ at $z=0$ obey the relations

$$\displaylines{\Psi_1^*(x,y,0,t)=\Psi_4(x,y,0,t),\cr\hfill \Psi_2^*(x,y,0,t)=\Psi_3(x,y,0,t)=0,\hfill}$$
that together with Eqs.(55)-(57) leads to the result
$\Sigma_x(x,y,0,t)=\Sigma_y(x,y,0,t)=\Sigma_z(x,y,0,t)=0$.

The analysis of expressions (46)-(48) and (57) shows that in
$x,z$-plane the $z$-component of spin density is an antisymmetric
function of $z$ that is connected with discussed above symmetry of
probability density with respect to the replacement $z\rightarrow
-z$.

\begin{figure}
  \centering
  \includegraphics[width=54mm]{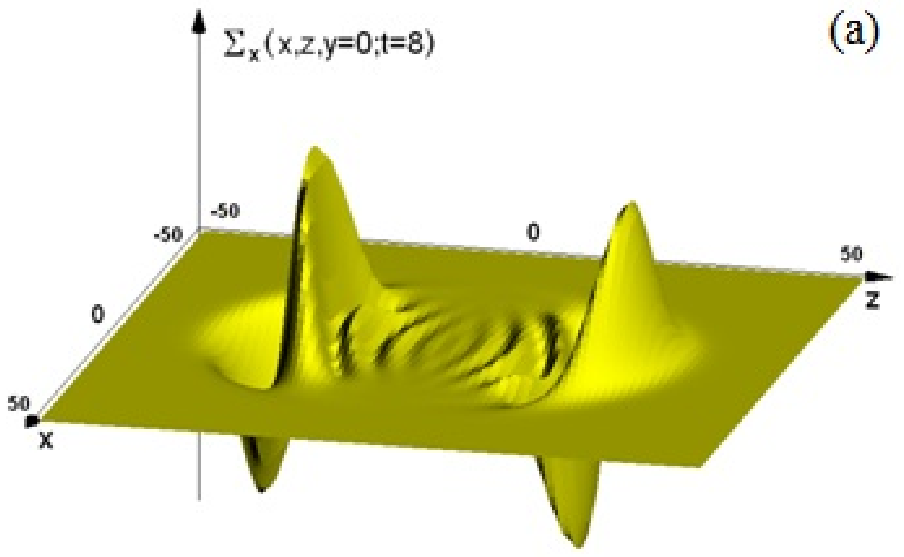}
  \includegraphics[width=54mm]{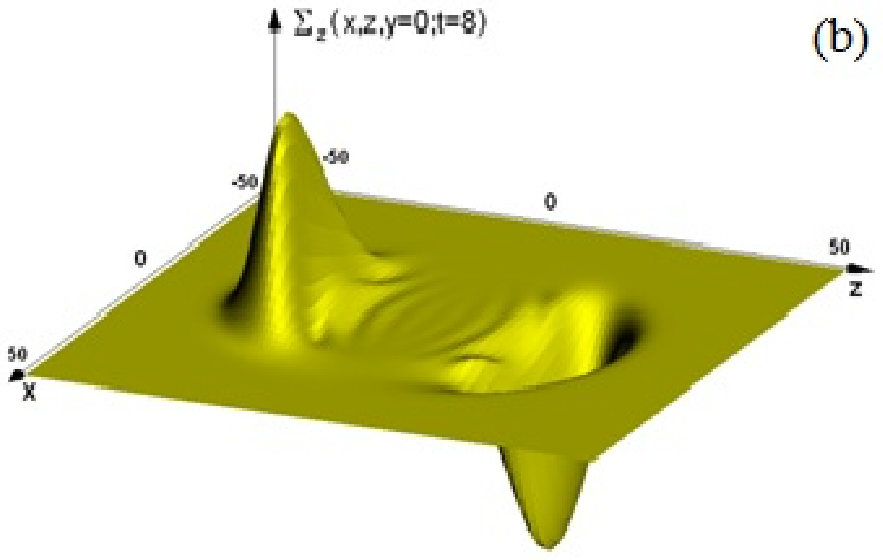}
\caption{(Color online). The distributions of the components of
spin density $\Sigma_x(x,0,z,t)$, $\Sigma_y(x,0,z,t)$ for initial
Gaussian packet, Eqs. (42), (34) with $k_0=0$ and $\Delta=d=3.64$
at time $t=8$.}\label{Fig6}
\end{figure}

Fig. 7 illustrates the distributions of $\Sigma_x(x,0,z,t)$ and
$\Sigma_z(x,0,z,t)$ for Gaussian wave packet with $k_0=1$ and
$\Delta=d=3.64$, at $t=8$.

By inserting Eqs.(43)-(45) into Eq.(54b) we find for the average
components of spin

$$\bar{\Sigma}_x(t)=-\int\frac{|f(\vec{p})|^2}{\lambda_{\vec{p}}^2}mc^3p_3(1-\cos\frac{2\lambda_{\vec{p}}t}{\hbar})d\vec{p},\eqno(62)$$

$$\bar{\Sigma}_y(t)=\int\frac{|f(\vec{p})|^2}{\lambda_{\vec{p}}}cp_3
\sin\frac{2\lambda_{\vec{p}}t}{\hbar}d\vec{p}.\eqno(63)$$

$$\displaylines{\bar{\Sigma}_z(t)=\int\frac{|f(\vec{p})|^2}{\lambda_{\vec{p}}}\Bigg[-cp_2\sin\frac{2\lambda_{\vec{p}}t}{\hbar}+
\frac{mc^3p_1}{\lambda_{\vec{p}}}\times \cr\hfill
\times(1-\cos\frac{2\lambda_{\vec{p}}t}{\hbar})\Bigg]d\vec{p}.\hfill\llap{(64)}\cr}$$
Obviously for a symmetric wave function $f(\vec{p})=f(-\vec{p})$
(that means $k_0=0$) these expressions give
$\bar{\Sigma}_x=\bar{\Sigma}_y=\bar{\Sigma}_z=0$. Note that
$\bar{\Sigma}_z$ remains to be equal to zero also for
$\bar{p}_z=\hbar k_0\neq 0$. It may be shown that in this case
Eqs.(62),(63) describe the spin "precession" (in $x,y$-plane about
vector $\vec{k_0}$) which has a transient character. Such
phenomenon for a hole system, described by Luttinger model was
discussed in the recent work of authors.\cite{DMF}

\section{Summary}

 In this work we have studied the quantum dynamics of
relativistic particles represented by three-dimensional Gaussian
wave packets with different initial spin polarizations, described
by the Dirac equation. The analysis of the general symmetry
properties of solutions of one-particle Dirac equation allows to
predict the direction of average electron velocity as well as the
direction of trembling motion. In particular, the evolution of
spherically and cylindrically symmetric initial Gaussian wave
packet demonstrates that the wave packet with initial
polarization, which is determined by bispinor $\pmatrix{1 & 0 & 1
& 0}^T$ has cylindrical symmetry at all times $t>0$, but the wave
packet with initial polarization $\pmatrix{1 & 0 & 0 & 1}^T$ loses
its cylindrical symmetry at time $t>0$. The influence of the
symmetry of initial wave packet on the distribution of spin
densities is analyzed.

\section*{Appendix A}

The "leap-frog" algorithm\cite{S} is applied in a spatial grid of
bin-sizes $\Delta x$, $\Delta y$, $\Delta z$ and with time step
$\Delta t$:

$$\Psi(\vec{r},t+2\Delta t)=\Psi(\vec{r},t)-2i\Delta t\hat{H}(\vec{r})\Psi(\vec{r},t+\Delta t),\eqno(A.1)$$
The spatial derivatives are computed symmetrically. Reflecting
boundary conditions are applied on a very large grid (running
stops before reflections occur if necessary). We use the norm as
the stability measure of the algorithm (1). In accordance with von
Neumann stability analysis\cite{GKO} (for large component plane
waves) the stability region ($d$-spatial grid bin, $\Delta t$-time
step) is:

$$d^4\Bigg(1-(\Delta t)^2\Bigg)-2(d\Delta t)^2-4(\Delta t)^2>0,\eqno(A.2)$$

Thus, for a single precision calculation the loss of norm can be
kept within $10^{-7}$-$10^{-6}$ in a $10^3$ time step run. It
should be noted that in the case of {\it Zitterbewegung}, i.e. of
the spatial oscillation of the wave packet, one more condition
have to be imposed to the lattice sizes:

$$\Delta x\sim \Delta y\sim \Delta z<\frac{\hbar}{mc},\eqno(A.3)$$
The conditions (A.2) and (A.3) were fulfilled in all our
calculations.

\section*{Appendix B. Drift velocity for the arbitrary initial wave function}

As was shown in previous investigations the motion of the  Dirac
wave packet center does not obey classical relativistic
kinematics. In particular, besides the rapid oscillations (ZB) the
wave packet can drift with constant velocity although its average
momentum is zero. One can show that in this case the direction of
such motion coincides with the direction of initial average
velocity. In fact, for the second example considered in this work
(Eq. (42)) only $\bar{V}_x(t=0)=c$ and
$\bar{V}_y(t=0)=\bar{V}_z(t=0)=0$. Therefore such initial spin
polarization leads to the motion of the wave packet with constant
velocity along $x$ axis (see Eqs. (21),(26)). It is easy to see
that in the first example, Eq. (10), the wave packet at $t=0$ has
the velocity directed along $z$ axis, so the motion at $t>0$
occurs in this direction.

Let now find the drift velocity of particle for the case of
arbitrary initial polarization

$$\Psi(\vec{p},t)=Af(\vec{p})\pmatrix{\varphi_1 \cr \varphi_2 \cr \varphi_3 \cr \varphi_4},\eqno(B.1)$$
where $\varphi_i$ are the complex coefficients,
$A=\frac{1}{\sqrt{\sum\limits_i |\varphi_i|^2}}$, and $f(\vec{p})$
is to be determined from the Fourier expansion of coordinate wave
function $F(\vec{r})$. (We do not suppose that $F(\vec{r})$ and
$f(\vec{p})$ have any symmetry).

At $t>0$ the wave function in momentum representation is

$$\displaylines{\Psi(\vec{p},t)=\Psi_+(\vec{p},t)+\Psi_-(\vec{p},t)=\cr\hfill\sum\limits_{i=1}^2 C_i(\vec{p})U_i(\vec{p})\rm e^{-i\lambda_{\vec{p}}t/\hbar}+
\sum\limits_{i=3}^4 C_i(\vec{p})U_i(\vec{p})\rm
e^{i\lambda_{\vec{p}}t/\hbar},~\hfill\llap{(B.2)}\cr}$$ Using the
Eqs (5),(6) and  (8) for free Dirac spinors we find from Eq. (B.2)

$$\displaylines{C_1=Af(\vec{p})N(\varphi_1+\gamma(p_1-ip_2)\varphi_4+\gamma p_3\varphi_3),\cr\hfill
C_2=Af(\vec{p})N(\varphi_2+\gamma(p_1+ip_2)\varphi_3-\gamma
p_3\varphi_4),\hfill\cr\hfill C_3=Af(\vec{p})N(-\gamma
p_3\varphi_1-\gamma(p_1-ip_2)\varphi_2+\varphi_3),\hfill\cr\hfill
C_4=Af(\vec{p})N(-\gamma(p_1+ip_2)\varphi_1+\gamma
p_3\varphi_2+\varphi_4).\hfill\llap{(B.3)}\cr}$$ The density of
$\mu$-component of velocity ($\mu=1,2,3$) in the momentum space is

$$V_\mu(\vec{p},t)=\Psi^+(\vec{p},t)\hat{V}_\mu\Psi(\vec{p},t),~~\hat{V}_\mu=c\alpha_\mu.\eqno(B.4)$$
Obviously the time-independent part $V_{\mu 0}$ of $V_\mu
(\vec{p},t)$ is defined as

$$V_{\mu 0}(\vec{p})=\Psi_+^+(\vec{p},t)\hat{V}_\mu\Psi_+(\vec{p},t)+
\Psi_-^+(\vec{p},t)\hat{V}_\mu\Psi_-(\vec{p},t),\eqno(B.5)$$
where $\Psi_+(\vec{p},t)$  and $\Psi_-(\vec{p},t)$ corresponds to
the contribution of positive and negative energy into Eq.(B.2).
One may check that

$$(V_\mu)_{ij}=U_i\hat{V}_\mu U_j=\Bigg\{{\frac{cp_\mu}{\lambda_{\vec{p}}}\delta_{ij},~i,j=1,2\atop
-\frac{cp_\mu}{\lambda_{\vec{p}}}\delta_{ij},~i,j=3,4}.\eqno(B.6)$$
So, using Eqs. (B.2), (B.5) and (B.6) we obtain

$$\bar{V}_{\mu 0}=\int \frac{c^2p_\mu}{\lambda_{\vec{p}}^2}(|C_1|^2+|C_2|^2-|C_3|^2-|C_4|^2)d\vec{p}.\eqno(B.7)$$
Substituting the expression for the coefficients $C_i$, Eq.(B.3)
into the Eq.(B.7), we find the constant velocity of wave packet
center

$$\displaylines{\bar{V}_{\mu 0}=\frac{c}{\sum\limits_{i=1}^4|\varphi_i|^2}\int |f(\vec{p})|^2\frac{c^2p_\mu}{\lambda_{\vec{p}}^2}\Bigg[mc(|\varphi_1|^2+|\varphi_2|^2-
|\varphi_3|^2-|\varphi_4|^2)\cr\hfill+2p_1{\rm
Re}(\varphi_1^*\varphi_4+\varphi_3^*\varphi_2)+ 2p_2{\rm
Im}(\varphi_1^*\varphi_4+\varphi_3^*\varphi_2)+\hfill\cr\hfill
+2p_3{\rm
Re}(\varphi_1^*\varphi_3-\varphi_2^*\varphi_4)\Bigg].\hfill\llap{(B.8)}\cr}$$
It is convenient to represent this expression using the initial
components of velocity $\bar{V}_\mu (0)$, so that

$$\displaylines{\bar{V}_{\mu 0}=\bar{V}_\mu (0)\int
|f(\vec{p})|^2\frac{p_\mu ^2}{\lambda_{\vec{p}}^2}d\vec{p}+
mc^4\int |f(\vec{p})|^2\frac{p_\mu}{\lambda_{\vec{p}}^2}
\times\cr\hfill\times \frac{(|\varphi_1|^2+|\varphi_2|^2-
|\varphi_3|^2-|\varphi_4|^2)}{\sum\limits_{i=1}^4|\varphi_i|^2}d\vec{p}.\hfill\llap{(B.9)}\cr}$$
(Note that in this equation there is no summation over $\mu$). Let
the initial Gaussian wave packet be spherically symmetric, i.e.
$f(\vec{p})$ is determined by Eq.(39) with $d=\Delta$. Then if the
average momentum $\hbar k_0=0$ the second term in Eq.(B.9) equals
to zero and the value of integral in the first term does not
depend on  $\mu$. Thus the direction of the constant velocity of
wave packet coincides with the initial one. In the case $k_0d>>1$
as it follows from Eq.(B.9) $\bar{V}_{z0}>>\bar{V}_{x0}$,
$\bar{V}_{y0}$ and the asymptotic direction of the average
velocity is along $z$-axis, i.e. along the average momentum
$\bar{p}_z=\hbar k_0$. A similar result was obtained in our
previous work\cite{MDF}, concerning the propagation of the wave
packet in graphene.

Note that Eq.(A.8) is valid also for the most general initial wave
packet of the form

$$\Psi(\vec{p},0)=(\Psi_1(\vec{p}),\Psi_2(\vec{p}),\Psi_3(\vec{p}),\Psi_4(\vec{p}))^T,$$
if we rename
$\Psi_i(\vec{p})=\frac{f(\vec{p})\varphi_i}{\sqrt{\sum|\varphi_i|^2}}$
in Eq.(B.8) . It is not difficult to check that the expressions
for the constant components of velocity obtained earlier for the
examples Eq.(24) and Eq.(42) follow also from the general equation
(B.8).

\section*{Acknowledgments}
This work was supported by the Program of the Russian Ministry of
Education and Science "Development of scientific potential of High
education" (Project No. 2.1.1.2686) and Grant of Russian
Foundation for Basic Research (No. 09-02-01241-a), and by the
President of RF Grant for Young Researchers MK-1652.2009.2.

\end{document}